\documentclass[final,5p,twocolumn,authoryear]{elsarticle}

\usepackage{amssymb}

\usepackage[colorlinks,linkcolor=blue,citecolor=blue,urlcolor=blue]{hyperref}

\usepackage{color}
\newcommand\AND{\wedge}
\newcommand\OR{\vee}
\newcommand\NOT{\neg}

\journal{Progress in Biophysics \& Molecular Biology}

\begin{document}

\begin{frontmatter}

\title{Integration of cellular signals in chattering environments}

\author[Gaia]{P.~Ru\'{e}}
\ead{pau.rue@upc.edu}
\author[Gaia]{N.~Domedel-Puig}
\ead{nuria.domedel@upc.edu}
\author[Gaia]{J.~Garcia-Ojalvo\corref{cor}}
\cortext[cor]{Corresponding author}
\ead{jordi.g.ojalvo@upc.edu}
\author[Gaia]{A.~J.~Pons}
\ead{a.pons@upc.edu}
\address[Gaia]{Departament de F\'{i}sica i Enginyeria Nuclear,
Universitat Polit\`{e}cnica de Catalunya\\
Edifici GAIA,
Rambla Sant Nebridi s/n, 
08222 Terrassa, Spain}

\begin{abstract}

Cells are constantly exposed to fluctuating environmental conditions. 
External signals are sensed, processed and integrated by cellular signal transduction networks,
which translate input signals into specific cellular responses by means of
biochemical reactions. 
These networks have a complex nature, and we are still far from having a complete characterization of 
the process through which they integrate information, specially given the noisy environment in
which that information is embedded. Guided by the many instances of constructive influences
of noise that have been reported in the physical sciences in the last decades,
here we explore how multiple signals are integrated in an eukaryotic cell in the presence of
background noise, or chatter. To that end, we use a Boolean model of a typical human signal
transduction network. Despite its complexity, we find that the network is able to display simple
patterns of signal integration.
Furthermore, our computational analysis shows that these integration patterns depend
on the levels of fluctuating background activity carried by other cell
inputs. Taken together, our results indicate that signal integration is sensitive to environmental 
fluctuations, and that this background noise effectively determines the information
integration capabilities of the cell.

\end{abstract}


\end{frontmatter}

\section{Introduction} \label{Introduction}

Living organisms are not passive objects which filter out blindly the modifications of the environment in which they live. 
On the contrary, both unicellular and multicellular organisms are biochemical systems that receive information from 
different environmental sources,  process this information, and consequently make decisions. These actions allow them to adapt 
their behavior to the information received. 
The existence of different input signals thus explains why cells with identical genetic content may sometimes behave differently.
The net effect of this information-relay process
is the activation or inhibition of key regulatory elements that commit the cell to perform specific tasks, such as taking nutrients, 
secreting chemicals, proliferating, or communicating with other cells in a tissue. 
In this way, the cellular signaling machinery \citep{Kholodenko:2006p1312} acts as a decision-making device that maps the large number of 
inputs signals to a reduced set of behaviors characterized by specific outputs \citep{helikar-2008}.

\begin{figure}[htbp]
\begin{center}
\includegraphics[width=0.41\textwidth]{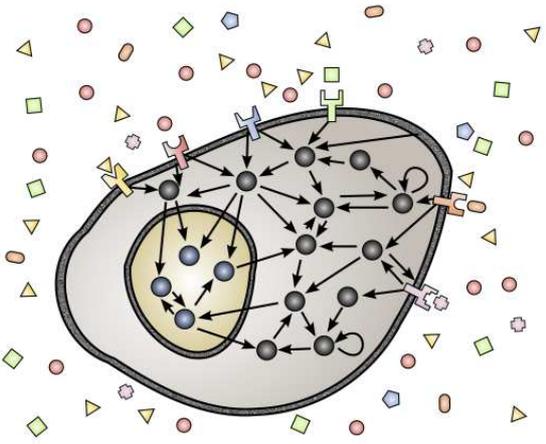}
\caption{Cells receive information from different sources coming from their environment, process this complex information through a complex web of protein interactions, and respond accordingly. 
\label{fig:CellDiagram}}
\end{center}
\end{figure}

All these physiological processes are implemented in the cell by signal transduction
networks composed of large numbers of interacting proteins. 
Many of these networks receive information about conditions outside the cell, 
which activate receptor proteins located at the membrane (see Fig.~\ref{fig:CellDiagram}). 
The activation of these receptors triggers a cascade of downstream biochemical reactions 
that ultimately modify the state of the cell, by controlling both DNA expression and protein levels.
Hence, signaling pathways integrate the signals coming from the environment, and define the response of the cell 
to the particular external situation in which it is living. 
These signaling routes have traditionally been studied as isolated cascades which do not interact with each other very strongly.
However, it is well known that the different signaling pathways are interconnected, and recent
studies have shown that this feature cannot be neglected 
\citep{Kaizu:2010p1124,Al-Shyoukh-2011}. In other words, one ought to refer to signaling networks,
rather than signaling pathways.
The fact that cells process information using an entire network of protein interactions means that the dynamical behavior 
can be much richer than the one obtained if signaling paths were uncoupled \citep{Bardwell:2007uq}.

These complex signal transduction processes must work reliably even in the presence of a substantial
amount of environmental noise. Indeed, the level of extracellular molecules representing a given
external signal is likely to fluctuate in time. More fundamentally, non-saturating levels of external signals
will induce stochastic activation of their receptors, leading to random fluctuations that are transmitted down
the signaling network and are bound to interfere with its information-processing tasks. The question
then arises, whether signaling networks have evolved to filter out the unavoidable levels of noise
existing in the extracellular milieu \citep{arias2006filtering}, or to use it in a constructive manner
\citep{steuer2003constructive,Turcotte:2008uq,Ullner:2009vn,Rue:2011kx}.

In fact, a large body of evidence has been gathered in the last decades within the fields of
statistical and nonlinear physics, showing that random fluctuations can be a source of order
(both temporal and spatial) in nonlinear systems \citep{gammaitoni1998stochastic,sagues2007spatiotemporal}.
A well-established example of such a constructive influence of noise is {\em stochastic
resonance}, a process through which an information-carrying signal (periodic or non-periodic) is
amplified by an optimal amount of noise acting on the system (internally or externally)
\citep{Wiesenfeld:1995p1219,moss1995benefits}. This
mechanism was initially proposed by \citet{benzi1981mechanism} as an explanation of the
occurrence of ice ages in earth's
climate (where the signal to be amplified corresponds to periodic variations in the amount of
energy received by the earth from the sun, and the noise arises from all other contributions to
climate fluctuations). After successful experimental demonstrations in lasers 
\citep{mcnamara1988observation} and in chemical systems \citep{kadar1998noise},
interest in this phenomenon grew and extended to biological systems,
namely sensory neurons, where the phenomenon was amply demonstrated
\citep{douglass1993noise,collins1995stochastic,gluckman1996stochastic,russell1999use,mori2002noise}.

Stochastic resonance arises in neuronal systems due to their excitable character
\citep{lindner2004effects}. Specifically, the noise-induced enhancement originates from the fact that
neurons are threshold
devices, and noise helps a time-varying (or space-varying) signal to surpass this threshold in a
manner that is correlated with the signal itself \citep{simonotto1997visual,stocks2000suprathreshold}.
This mechanism can act even if there is no external input signal acting upon the system, with
noise extracting instead an internal temporal (or spatial) scale existing within the system's dynamics.
This latter mechanism is called {\em stochastic coherence} \citep{zaikin2003doubly,carrillo2004spatial},
or more commonly {\em coherence resonance} \citep{pikovsky1997coherence,neiman1997coherence},
and has been suggested to exist in genetic oscillators \citep{vilar2002mechanisms}.

There are other examples of the constructive effects of noise beyond stochastic resonance and
stochastic coherence. Most notably, noise has been seen to induce transitions, both in time
\citep{hosthemke1984noise} and in space (in that case one speaks of noise-induced {\em phase}
transitions) \citep{garcia1999noise}. In those situations, the statistical properties of the system
change when noise is varied beyond a certain critical threshold, so that new states appear
as a result of random fluctuations. Experimental preliminar evidence of this type of behavior
in genetic systems has been presented by \citet{blake2003noise}, but there is still much work
to do in this direction.

In the context of cellular signaling, it is appealing to think that environmental noise might be
used by the cell in a constructive manner, similarly to what happens in stochastic resonance,
to detect and amplify information-carrying signals. Preliminary theoretical evidence in this
direction has been recently presented \citep{Domedel-Puig:2011fk}. However, the implication
of this type of mechanism on signal integration processes is still an open question. In fact,
despite the effort made during decades, signal integration is far from being completely understood. 
One of the reasons behind this problem is that a detailed theoretical description of a cell is proving difficult, 
because cells are extremely complex objects. The chemistry of cells operates at multiple levels,
spanning for instance signaling interactions, 
genetic expression, and metabolism, and each of these levels of operation is mediated
by many chemical species that interact 
in a strongly nonlinear manner, forming highly complex dynamical webs.  

In this paper, we discuss work that moves away from classical signaling studies that focus on only one ligand,
and examine instead the effects of multiple ligands upon the same cell, in the presence
of external noise. 
In particular, we consider whether the ability of cells to integrate signals from one or more sources of information is altered when
other input signals provide a chattering environment.
To study this problem, we consider a dynamic description of the signaling network of a typical human cell --a fibroblast-- 
in terms of a previously derived Boolean model \citep{helikar-2008}. 
Boolean models describe the activity of the different proteins involved in the network in a binary way,
as being ON or OFF, and are iterated in order to update the protein activity states according to
logical rules acting in each node. Even though there is no well defined time scale in this type of model,
Boolean models have been shown to reproduce ordered sequences of cellular events
\citep{li2004yeast,davidich-2008,bornholdt-2008}.
The model we use was constructed manually by \citet{helikar-2008} using the large body of scientific literature
available for this type of cell,
and it represents one of the largest signaling network models currently available.
Despite representing a simplified view of the real network, it offers a good balance between a realistic
description and a model simple enough 
to be studied theoretically. Indeed, this model was shown to reproduce experimental results satisfactorily \citep{helikar-2008}.
Using this model, we explore systematically, by means of numerical simulations, the effect of 
a chattering environment in the integration of input signals.

\section{Materials and methods}
\label{methods}

As described above, we
use a Boolean model of signal transduction to study signal integration in varying environments.
This choice is appropriate for our purposes, since it defines the level of activity of the molecules that compose the signaling network. Boolean networks (BN hereafter)
were introduced by \citet{kauffman-1969} as sets of dynamical units, called \emph{nodes}, 
which are connected with each other by interaction \emph{edges}. 
The state of every node $i$ at each discrete time $t$, $x_i(t)$, is either ON ($x_i(t)=1$) or OFF ($x_i(t)=0$),
and all nodes are updated according to a set of node-specific logic rules  that define the dynamics of the system.  
For signaling networks, the state of each node corresponds to the level of activity of a specific protein, and edges represent their interactions.
We will consider two types of nodes, determined by their definition:
\emph{input} nodes and \emph{internal} nodes. 
On the one hand, the state of \emph{input} nodes is independent from the state of other nodes of the network and determined externally. 
These nodes represent signals outside the cell (i.e. the external environment), and may evolve deterministically (for instance, 
being maintained to a fixed value of activity or oscillating at a specific periodicity) or stochastically.  
Here we set these states using random value sequences.
On the other hand, every \emph{internal} node $i$ has an associated logic rule, $f_i$, which determines the new state $x_i(t)$ at time $t$ 
from the states of its $k_i$ incident nodes at time $t-1$ (note that $k$ may include the node $i$ itself):
\begin{equation}
    x_i(t) = f_i(x_i^1(t-1), \ldots, x_i^{k_i}(t-1)).
\end{equation}

\begin{figure}[htbp]
\begin{center}
\includegraphics[width=0.4\textwidth]{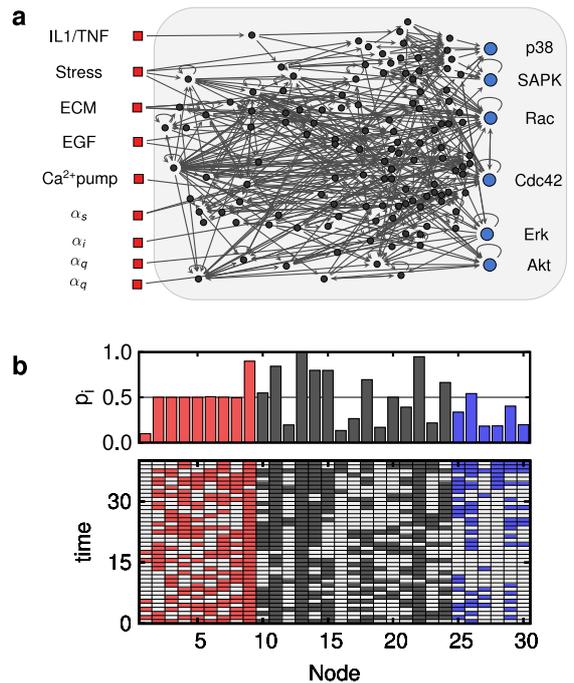}
\label{fig:Network}
\caption{(a) Fibroblast signaling network. It is formed by 9 input nodes (red squares) connected to 130 internal nodes (circles). Six of this internal nodes (blue circles) are associated with specific cellular responses, and thus
are considered to be output nodes. The 542 directional connections between interacting nodes are represented as arrows. (b) An example of the network dynamics (input nodes in red, 15 internal nodes in gray and output nodes in blue). Time evolution of states of some network nodes (bottom panel); OFF (0) states are
shown in white and ON (1) states are shown in colors. The temporal average of each node activity is shown in the top panel.}
\end{center}
\end{figure}
The fibroblast signal transduction network analyzed in this paper \citep{helikar-2008}
consists of nine input nodes and 130 internal nodes interconnected 
by a dense web of interactions (see Fig.~\ref{fig:Network}a).
The input layer contains cytokines (IL1/TNF-$\alpha$), growth factors (EGF),
extracellular matrix components (ECM),
oxidative stress, a calcium pump and different G protein-coupled receptor ligands.
\citet{helikar-2008}  chose six of the 130 internal nodes to be \emph{outputs} of the network 
(the proteins Akt, Erk, Rac, Cdc42, SAPK and p38). 
The motivation for choosing these six output species was their role in regulating specific cellular processes:
apoptosis (Akt), gene transcription (Erk), cytoskeletal regulation (Rac and Cdc42), and cellular stress (SAPK and p38).

The activity of every input node $i$ has been implemented as a stochastic sequence in which each state (0 or 1)
derives from an independent draw of a Bernoulli distribution with success probability equal to $q_i$ (where $q_i$
is held constant within each run). 
In this way, the parameter $q_i$ is what we define as the \emph{chatter level} for node $i$.
The produced sequences of states, $x_i(t)$, show an intrinsic variability, 
which is maximal at $q_i=0.5$, and decreases monotonically when $q_i$ approaches to 0 and 1.
Here we calculate the Bernoulli draws with a standard Mersenne twister pseudo-random number generator~\citep{matsumoto-1998}.

We are interested in the average activity of each node $j$ over time and over multiple realizations
(which might correspond to different cells in a population):
\begin{equation}
p_j= \langle x_j\rangle =
\frac{1}{N_{\rm cells}T}\sum^{N_{\rm cells}}_{{\rm cell}=1}\sum^{T}_{t=0} x_{j,{\rm cell}}(t).
\end{equation}
Note that, for input nodes and sufficiently long dynamical evolutions, $q_i=p_i$.
To illustrate this procedure, the upper panel in Fig.~\ref{fig:Network}b shows the average activity $p_i$ of 30 network nodes
(including inputs, outputs, and internal nodes) and the corresponding sequence of states is shown in the lower panel.
The time evolution of the system has been obtained using a custom-made Python implementation
of a BN simulator, available at \href{http://code.google.com/p/bnsim}{http://code.google.com/p/bnsim}.

\section{Results and discussion}
\label{Results}

\subsection{Network dynamics and chatter effects on information flow}

Despite the simplicity of the Boolean aproximation, the network model described in the preceding Section
shows interesting behavior in terms of the dynamical integration of input signals.
For instance, when the system receives constant inputs it falls in periodic orbits within a distribution of transient times peaked at one specific length, but with a long-tailed decay in the form of a power law \citep{rue:045110}. This characteristic transient scale, which may be relevant for the operational response of the cell to changes in the environment, is maintained in the presence of noisy environments characterized by a chatter level, $q_i=q$. Thus, even though chatter destroys the periodic orbits, it maintains the fixed transient response at the population level (considering the average activity of different realizations of the dynamics).
Besides, the average activity of the outputs also depends on the chatter level \citep{Domedel-Puig:2011fk}.
This property is a result of the structure of the biologically network studied here (its topology and its logic rules) because it is not observed in randomized versions of the network
with the same topological properties but with different logic rules for the nodes. In other words,
the biologically realistic network responds in a nontrivial manner to input signals with constant level of background chatter, and it is the distribution of logic rules of the network nodes
and their particular placement within the network what determines the responsiveness.

Background chatter also enhances the network response to periodic stimulation \citep{Domedel-Puig:2011fk}.
Again, this dynamical behavior is not shown for randomized versions of the network. This dynamical response
is related with the characteristic time response of the network~\citep{rue:045110}. In fact, the network responds differently to different stimulation frequencies for specific input nodes. The average activity of the
output nodes shows low-pass, high-pass or band-pass frequency filtering responses depending on the
node. Indeed, noise intensity, as introduced by chatter levels in the input signals,  seems to be a key factor in determining the paths for which information flows from periodically stimulated input nodes to output nodes~\citep{Domedel-Puig:2011fk}.

Altogether, these results reveal that the dynamical response to external modulation is, at the same time, versatile and robust to noise. The extremely rich dynamical behavior observed in this model, consistent with the diversity of responses observed in cellular organisms, is the result  of the complex processing information ability of the network. The ability to combine the information from different input sources to produce a coherent output response is possible only if the different signaling pathways can interact within the network. In fact, \citet{Natarajan-2006} introduced the term ``interaction agent'' to refer to the network circuits that couple different signaling pathways. We are interested in understanding how noise affects pathway interaction and thus
signal integration.

\subsection{Integration of a signal with background chatter}

The above-described ability of chatter to select the information route going from the (periodically driven)
input node to the output nodes stresses once more that the complex network is not a passive object that
merely transmits information, but it also processes this information, modifying it, when going from the emitting (input) node to the receiver (output) node. This phenomenology leads us to ask if the background chatter level
$q$ modifies the activity of the network not only when one node is perturbed periodically,
but also when this node is maintained at a specific (different) chatter level $q_1$. In other words,
we ask if the routing ability of chatter revealed in the dynamical context of one periodic input
also exists in a more static situation, when that node is maintained at a specific chatter level $q_1$,
different from the fixed background chatter level $q$ that affects all other input nodes.

In that direction, we now systematically study the average activity of the output nodes when changing the specific chatter level, $q_1$, of a given input node,
under different values of the global background chatter, $q$. That is, we scan the model in two dimensions for each input $i$:
for $q_i$ ranging from 0 to 1 (in 0.10 steps), and a common $q$ equal for the remaining 8 inputs, also ranging from 0 to 1.
We compute the average activity of the output nodes at steady state over different realizations, 
arising from 200 different initial conditions and realizations of the stochastic variables.

\begin{figure*}[htbp]
\begin{center}
\includegraphics[width=0.8\textwidth]{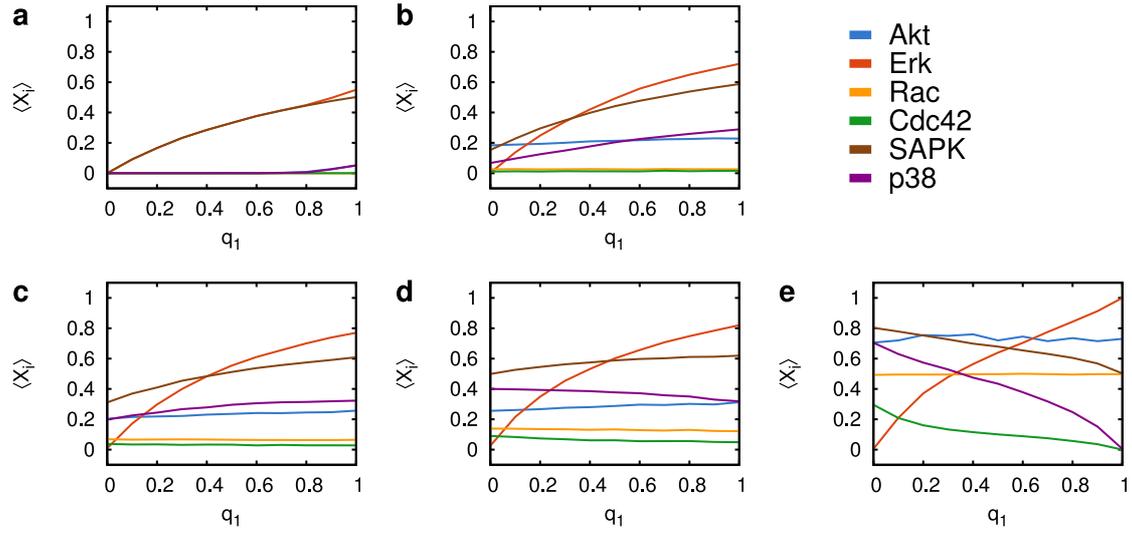} 
\caption{Average activity of the output nodes versus the chatter level, $q_1$, of the input node IL1/TNF-$\alpha$ for different values of the background chatter $q=0.0, 0.3, 0.5, 0.7, 1.0$. 
Averages are computed over 200 cell realizations lasting 300 time iterations each, after
dismissing a transient of 100 iterations.  
\label{fig:q1}
}
\end{center}
\end{figure*}

The quantitative behavior of the network outputs is specific for each of the considered inputs.
Furthermore, in most cases the dependence of the activity of output nodes on $q_i$ changes
when the common background chatter $q$ is varied.
This can be seen for instance in Fig.~\ref{fig:q1}, which for the sake of simplicity shows  
only the case when the chatter of the input node IL1/TNF-$\alpha$, $q_1$, is modified.
These results confirm the fact that the integration of one signal depends on the
rest of input signals. 
Different output nodes in Fig.~\ref{fig:q1} display different behaviors:
for instance, the monotonic increase of Erk with $q_1$ is basically unchanged when increasing the global background activity $q$. 
The opposite situation is shown for SAPK and p38, which swap their monotonically increasing dependence with $q_1$ at lower $q$
to a monotonically decreasing dependence at high $q$. These two examples highlight the dramatic influence of the background signaling
upon the processing of one specific ligand. 
In this case, a low activity of the background signals would enhance the activity of SAPK and p38 when IL1/TNF-$\alpha$ increases, 
but would inhibit their output activity with IL1/TNF-$\alpha$ when the background chatter is larger. 
The average activities of Akt, Rac and Cdc42 are independent of the degree of activity of IL1/TNF-$\alpha$, 
but grow when the global background chatter level increases. 
Thus, these output nodes are not sensitive to IL1/TNF-$\alpha$ but they are activated by the background chatter. 
Qualitatively similar changes with $q$ are observed for other inputs.

\subsection{Integration of two signals under background chatter}

\begin{figure*}
\begin{center}
\includegraphics[width=0.85\textwidth]{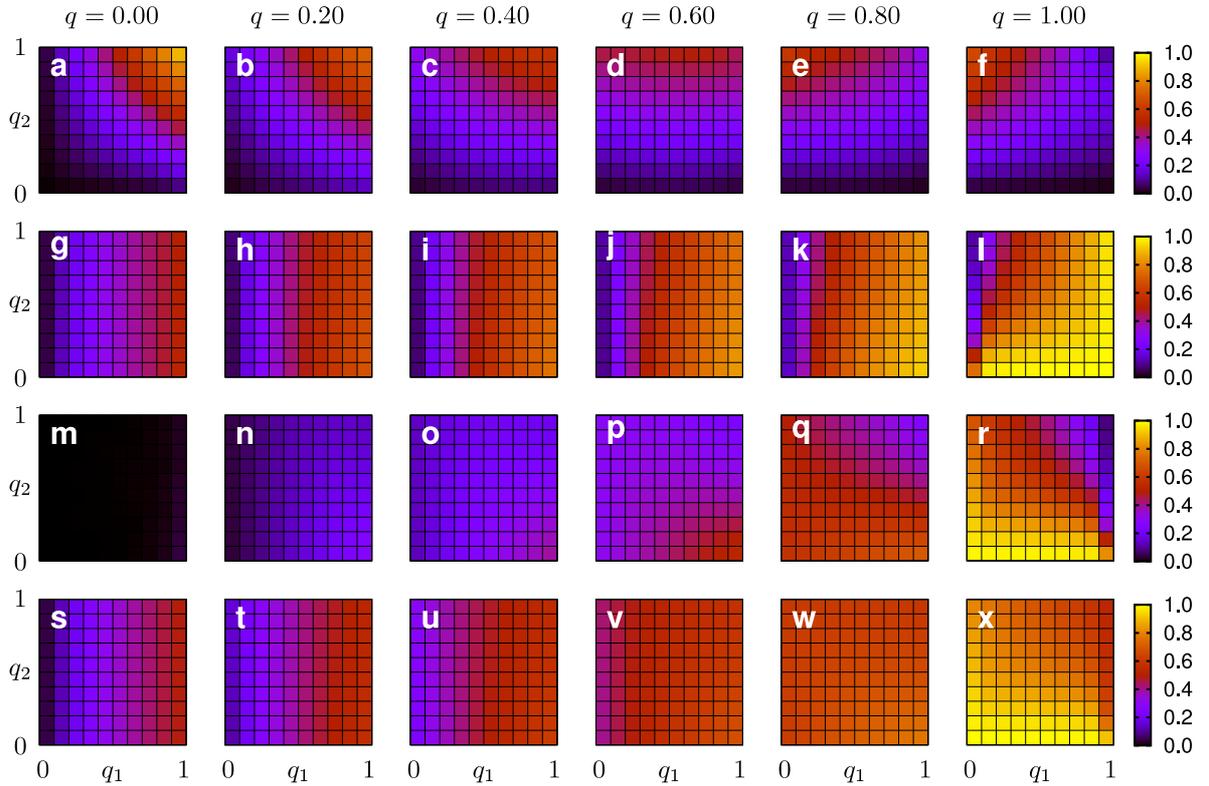}
\caption{(color online) (a) Integration maps for four combinations of two inputs and an output
for varying background chatter $q$.
Averages are computed, for each pair of $q_1,q_2$ values, over 200 cell realizations
of duration 300 iterations, and after dismissing an initial transient of 100 iterations.
The first input, measured by $q_1$, is in all cases IL1/TNF-$\alpha$. The second input ($q_2$)
and the output are, respectively: (a-f) EGF and p38, (g-l) ECM and Erk, (m-r) ECM and p38,
and (s-x) ECM and SAPK.
\label{fig:q1q2} }
\end{center}
\end{figure*}

The results of Fig.~\ref{fig:q1} support the idea that 
the integration of one single input
cannot be studied in isolation from the activity of the other inputs received by the cell. 
These results are consistent with the dependence of the average activity of some output nodes
on background chatter level,
and with the effect on the integration of periodic signals shown recently by \citet{Domedel-Puig:2011fk}. 
Continuing in this direction, 
we now examine the effect of background chatter when two different input nodes are activated with two different 
and specific chatter levels, $q_1$ and $q_2$, respectively. To that end, we perform a comprehensive set of numerical simulations 
combining all the possible pairs of inputs, with the values for the $q$ common to the remaining seven inputs ranging from 0 to 1,
and averaging over 200 different initial conditions and realizations of the random variables.
Four examples of the outcome of these simulations are presented 
in Fig.~\ref{fig:q1q2}, which shows the average activity (in color code) of the different output nodes 
when integrating the levels of various pairs of input nodes, for increasing background chatter levels,
as indicated. 
The output activity surfaces obtained in this way, or \emph{integration maps}, change as the background
chatter level increases. 
In particular, in the case of IL1/TNF-$\alpha$ and EGF as inputs [Figs.~\ref{fig:q1q2}(a-f)],
both inputs have to be ON at low background chatter
to activate p38, whereas at high background chatter EGF has
to be ON and IL1/TNF-$\alpha$ has to be OFF for p38 to be turned on.
In other cases the network goes from being sensitive only to input $q_1$ [Figs.~\ref{fig:q1q2}(g,s)])
to integrate the two inputs in different ways [Figs.~\ref{fig:q1q2}(l,x)]), as the
background chatter increases. The network can also transition from being completely insensitive
to both inputs [Figs.~\ref{fig:q1q2}(m)] to develop an integration capability in the presence of noise
[Figs.~\ref{fig:q1q2}(n-r)]. Other combinations of input pairs and outputs exhibit nontrivial changes
in their integration maps as chatter varies.

\subsection{The network integration logic changes with chatter}

\begin{figure}[htbp]
\begin{center}
\includegraphics[width=0.4\textwidth]{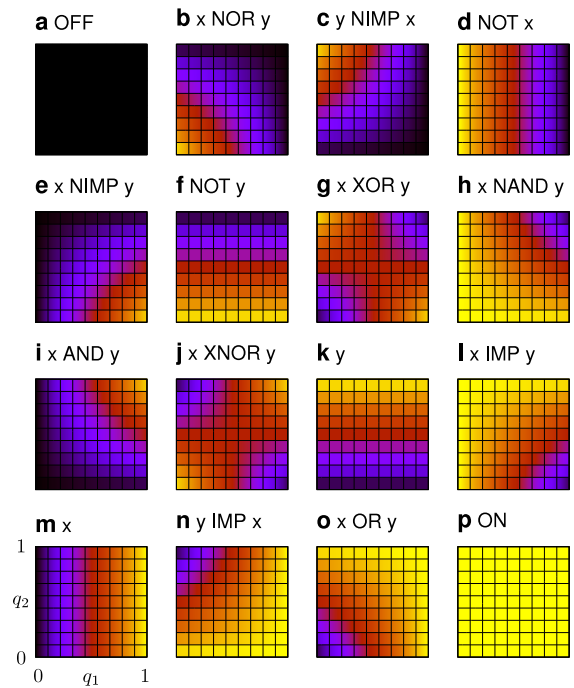}
\caption{Integration maps for the 16 different rules with two inputs.
The names of the logic gates are based on \citet{Mayo:2006fk}.
The color coding is the same as in Fig.~\protect\ref{fig:q1q2}.
\label{fig:LogicMaps}
}
\end{center}
\end{figure}

The transitions observed in Fig.~\ref{fig:q1q2} are very relevant for cellular behavior. 
Their existence suggests that the net effect of the network 
is to act as a switch mechanism which selects the type of $q_1$/$q_2$ integration as a function of the different background chatter levels. 
This behavior may be interpreted in terms of simple logic gates with two inputs.
Figure~\ref{fig:LogicMaps} shows the integration maps corresponding to all possible two-input
logic gates:
two of these (invariant ON and invariant OFF) are independent of the inputs, 
four of them are sensitive to only one of the inputs ('x', 'NOT x', 'y' and 'NOT y'),
and the rest (ten gates) integrate the information from both inputs in order to produce the output.
These integration maps can be computed analytically, assuming that the gates are supplied with
two independent and binary random processes, $x_1$ and $x_2$, with average values
$q_1$ and $q_2$. The result of this calculation is given in Table~\ref{Ta:LogicMaps}.
If we compare the surfaces shown, for instance, in Figs.~\ref{fig:q1q2}(a) and (f)
to the aforementioned 16 basic gates, we observe that those two limit cases are very similar
to the 'x AND y' and 'y NIMP x' gates. 
Hence, despite the complexity of the network under study, we observe that some output nodes
behave as functions of simple two-input logic gates. Importantly, the level of background
chatter selects the effective logic gate that will operate at any given condition.

\begin{table}[htbp]
\begin{center}
{\small
\begin{tabular}{c|c|c|c}
{\bf Rule}	& {\bf Expr. ($r_i$)}			& {\bf Prob. ($p$)}			& {\bf Name} \\
\hline
$r_0$			& 0					& $0$					& OFF \\
$r_1$			& $\NOT(x\OR y)$		& $1-q_1-q_2+q_1q_2$		& x NOR y \\
$r_2$			& $\NOT x\AND y$		& $q_2-q_1 q_2$			& y NIMP x \\
$r_3$			& $\NOT x$			& $1-q_1$					&  NOT x\\
$r_4$			& $ x\AND \NOT y$		& $q_1-q_1 q_2$			& x NIMP y\\
$r_5$			& $\NOT y$			& $1-q_2$					&  NOT y\\
$r_6$			& $x\AND\NOT y\OR\NOT x\AND y$		& $q_1+q_2-2q_1q_2$		& x XOR y \\
$r_7$			& $\NOT(x \AND y)$		& $1-q_1q_2$				& x NAND y\\
$r_8$			& $x \AND y$			& $q_1 q_2$				& x AND y\\
$r_9$			&$\NOT(x\AND\NOT y\OR\NOT x\AND y)$	& $1-q_1-q_2+2q_1q_2$		& x XNOR  y\\
$r_{10}$			& $y$				& $q_2$					& y  \\
$r_{11}$			& $\NOT x \OR y$		& $1-q_1+q_1 q_2$			& x IMP y \\
$r_{12}$			& $x$				& $q_1$					& x \\
$r_{13}$			& $x \OR\NOT y $		& $1-q_2+q_1 q_2$			& y IMP x\\
$r_{14}$			& $x \OR y$			& $q_1+q_2-q_1q_2$			& x OR y\\
$r_{15}$			& $1$				& $1$					& ON \\
\end{tabular}
}
\caption{Definition of the 16 integration rules with two inputs shown in Fig.~\protect\ref{fig:LogicMaps}.
\label{Ta:LogicMaps}}
\end{center}
\end{table}

\begin{figure}
\begin{center}
\includegraphics[width=0.47\textwidth]{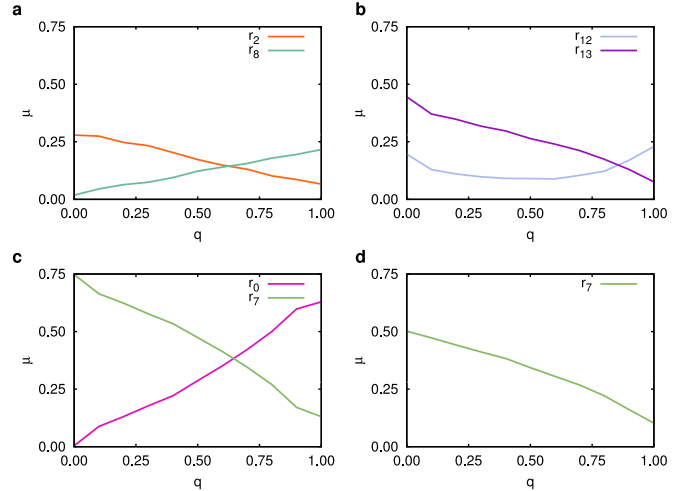}\\
\caption{Transition between simple logic gates, as measured by $\mu_i(q)$, for increasing background
chatter $q$, for the four cases of Fig.~\ref{fig:q1q2}. Only the logic gates for which the mean
absolute difference is smaller than a given threshold (here taken equal to 0.14 without loss
of generality) for any value of $q$ are shown.
The meanings of the different lines given in the legends
follow the notation of Table~\ref{Ta:LogicMaps}.
\label{fig:transition}}
\end{center}
\end{figure}

To compare the input-output transfer function of the network at different background chatter
levels, $f(q_1,q_2;q)$, 
with the simple integration patterns given by the two-input logic functions, $r(q_1,q_2)$, we use
the {\em mean absolute difference} measure, 
$\mu_i(q)$, defined as:
\begin{equation}\label{eq:MAD}
\mu_i(q) = \int\left| f(q_1,q_2;q)  - r_i(q_1, q_2) \right|\,dq_1\,dq_2,
\end{equation}
where $i$ runs over the 16 gates (see Table~\ref{Ta:LogicMaps}).
We consider all the combinations of pairs of input nodes, under the full range of background chatter values.
A diversity of behaviors is observed: in some cases, 
the behavior obtained from the dynamical evolution of the network cannot be attributed to any single logic gate. Instead, it displays
a combination of gates at some specific background chatter level
(see for instance all cases in Fig.~\ref{fig:q1q2} in which the responses 0 --black-- and 1 --yellow--
are absent). In other cases, 
such as the one corresponding to the top row in Fig.~\ref{fig:q1q2}, the situation is simpler and
a somewhat clearer transition between simple logical gates is observed 
as background chatter is modified. This transition is quantified in terms of $\mu_i(q)$ and shown in
Fig.~\ref{fig:transition}(a). Other, somewhat less clear, transitions between simple logic gates are
shown in Figs.~\ref{fig:transition}(b-d).

The transitions between simple binary logic gates are, to some extent, a generalization of the
results obtained for the integration between a single input and background chatter. The generalization to interactions obtained for an increasing number of input signals 
may define a hierarchy of simple behaviors which somehow classifies widely varying input signal
combinations into a set of simple biological responses~\citep{helikar-2008}. These results show that the combination of the topology of the network and the specific logic functions defined at each node or, in other words, the full set of protein-protein interactions of the signaling network, have to be carefully chosen. The network has to be constructed in such a way that this hierarchy of integration levels, which classify the output response of the cell, end up in a well-defined pattern of robust responses that are biologically useful.

\section{Conclusions} \label{Conclusions}

Given the many examples existing in the physical sciences of the constructive influence of random
fluctuations, it is appealing to think that signaling networks have been tailored by evolution to make
use of the unavoidable noise to which cells are subject, in particular from its environment. Here
we have addressed the issue of how background noise affects the signal integration capabilities of
the signaling network of a typical human cell, namely a fibroblast. Using a Boolean model of this
network, we have analyzed the integration of one or more signals for different levels of background
chatter affecting the rest of inputs of the network. We have observed that: (i) the response to an input signal
depends on the level of background chatter affecting all other input nodes,
(ii) even though the network is highly complex, the outputs respond in a very simple way to pairs of
interacting input signals, and (iii) variation of the chatter level causes transitions between these simple
pattern responses. 
This type of behavior has been observed experimentally \citep{Janes:2005fk,Hsueh2009}
and might suggest that the cell adapts its information processing machinery according to the environment.

\section*{Acknowledgements} \label{Acknowledgements}

This work has been financially supported by the Spanish Network of Multiple Sclerosis
(REEM, Instituto de Salud Carlos III), the Fundaci\'on Mutua Madrile\~na (Spain), the Ministerio de
Ciencia e Innovaci\'{o}n (project FIS2009-13360), and the Generalitat de Catalunya (project 2009SGR1168).
P.R. acknowledges financial support from the Generalitat de Catalunya.
J.G.O. also acknowledges financial support from the ICREA foundation.

\end{document}